\documentclass[a4paper]{jpconf}
\usepackage{graphicx}
\usepackage{JHEP_PR_dilasym}
\usepackage{atlasphysics}
\usepackage{subfigure}
\usepackage{caption}
\usepackage{lineno}
\begin{document}
\title{Measurement of the charge asymmetry in dileptonic decays of top quark
  pairs in \pp collisions at $\rts=7$\,\tev\ using the \atlas detector}

\author{C Deterre$^{1}$ and L Mijovi\'c$^{2,3}$ on behalf of the ATLAS Collaboration}

\address{$^{1}$DESY, Notkestrasse 85, 22607 Hamburg, Germany\\
         $^{2}$Irfu/SPP, CEA-Saclay, 91191 Gif-sur-Yvette Cedex, France\\
         $^{3}$Universit\"{a}t Bonn, Physikalisches Institut, Nussallee 12, 53115 Bonn, Germany}
                 
\ead{$^{1}$cecile.deterre@cern.ch, $^{2}$liza.mijovic@cern.ch}


\begin{abstract}
A measurement of the \ttbarw (\ttbar) charge asymmetry is presented 
using 2011 LHC data collected by the \atlas detector corresponding to an integrated luminosity of \lumi{} at a 
centre-of-mass energy of 7\,\tev. 
The analysis is performed in the dilepton channel, and two different observables are studied: \Acll{}, based on the identified charged
leptons, and \Actt, based on the reconstructed \ttbar\ final state.  
The asymmetries, measured to be \Acll~=~\Acllcomb\ and \Actt~=~\Acttcomb, 
are in agreement with the Standard Model predictions.
\end{abstract}

\section{Introduction}

This analysis~\cite{asym-ll-7TeV} uses a data set corresponding to an integrated luminosity of
\lumi{} of Large Hadron Collider (LHC)  proton--proton (\pp{}) collisions at a
centre-of-mass energy of 7\,\tev\ collected by the \atlas~\cite{atlas} detector. It is
performed in the dilepton channel of the \ttbar{} decay. The measured observables are the lepton-based charge asymmetry \Acll and the \ttbar{} charge asymmetry \Actt{}.
\Acll is defined as an asymmetry between positively and negatively charged leptons:
\begin{linenomath}
\begin{equation}
\label{eq:ac_lep}
\Acll = \frac{N(\Delta |\eta| >0) - N(\Delta |\eta| <0)}{N(\Delta |\eta|
      >0) + N(\Delta |\eta| <0)}, 
\end{equation}
\end{linenomath}
where $\Delta |\eta| = |\eta_{\ell^{+}}|-|\eta_{\ell^{-}}|$, $\eta_{\ell^{+}}$ ($\eta_{\ell^{-}}$) is the pseudorapidity\footnote{The pseudorapidity is defined in terms of the polar angle $\theta$ as $\eta=-\ln\tan(\theta/2)$.} 
of the positively (negatively) charged lepton and $N$ is the number of events
with positive or negative $\Delta |\eta|$.
The \Actt{} corresponds to the asymmetry in top and
antitop quark rapidities\footnote{The rapidity is defined as: 
$y = \frac{1}{2} \ln\frac{E + p_z}{E - p_z}$ where $E$ is the energy of the
particle and $p_z$ is the component of the momentum along the LHC beam axis.}:
\begin{linenomath}
\begin{equation}
\label{eq:ac}
\Actt = \frac{N(\Delta |y| >0) - N(\Delta |y| <0)}{N(\Delta |y|
  >0) + N(\Delta |y| <0)}, 
\end{equation}
\end{linenomath}
where $\Delta |y| = |y_{t}|-|y_{\bar{t}}|$, $y_{t}$ ($y_{\bar{t}}$) is the rapidity of the top (antitop) quark, and $N$ is the number of events with positive or
negative $\Delta |y|$.

In Standard Model (SM) \ttbar{} production, the asymmetry is absent at leading-order (LO) Quantum Chromodynamics (QCD) and is introduced by the next-to-leading-order (NLO) QCD contributions to the \ttbar{} differential
cross-sections, which are odd with respect to the exchange of $t$ and $\tbar$~\cite{Bernreuther:2012sx}.
The Tevatron experiments have reported deviations of closely related observables, the forward-backward asymmetries, from the SM predictions~\cite{CDF1,D01}. While an agreement with the SM has been obtained for a number of asymmetry measurements at the Tevatron and the LHC since then, the deviations still persist in recent results reported by the
CDF collaboration~\cite{CDF2}. These deviations and the sensitivity of the asymmetry to beyond 
Standard Model (BSM) effects motivate the measurement.

\section{Event selection}
\label{Sec:Esel} 
 
\begin{figure}
\centering
\subfigure[] {
\includegraphics[width=0.40\textwidth]{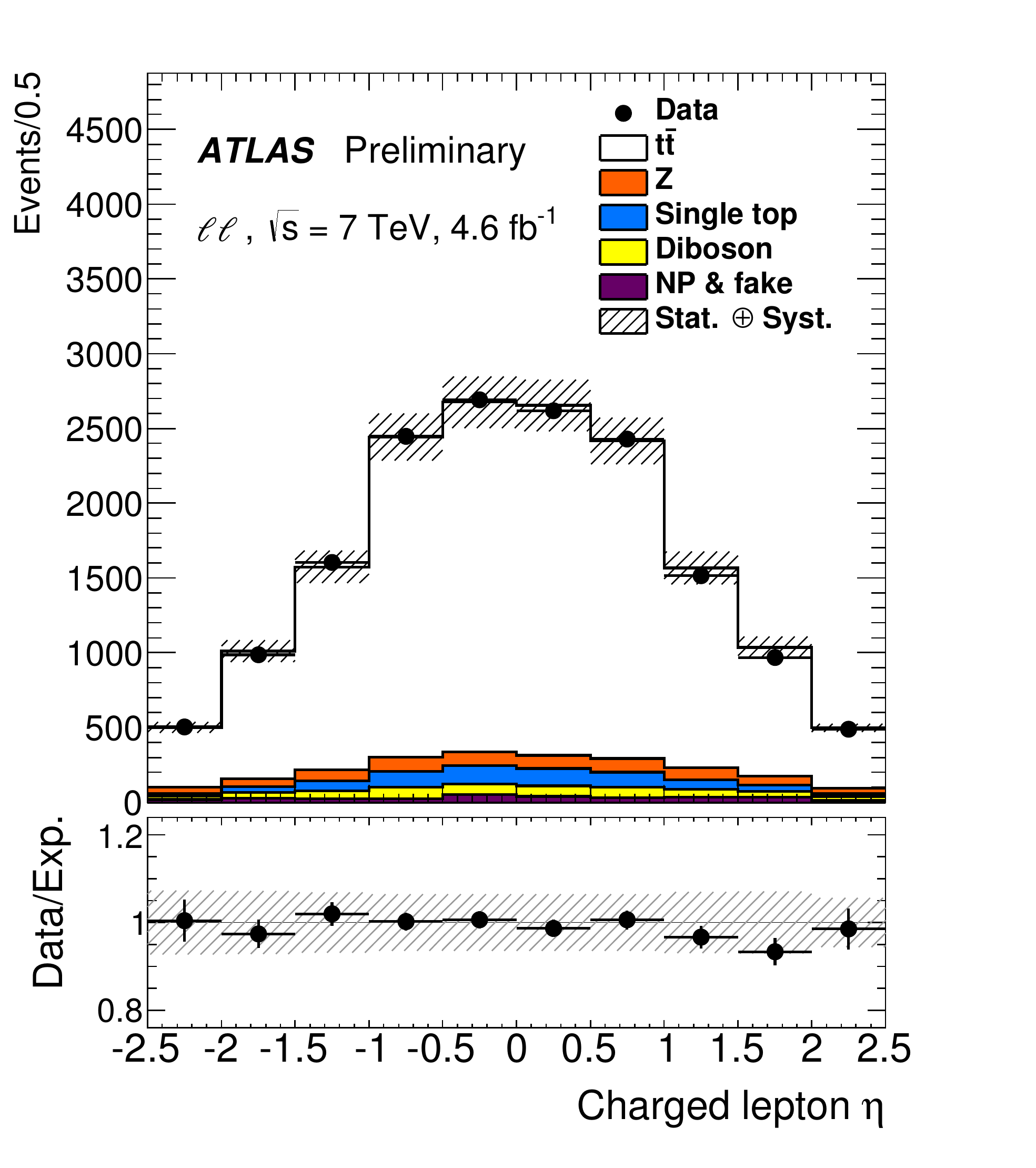}
}
\subfigure[] {
\includegraphics[width=0.40\textwidth]{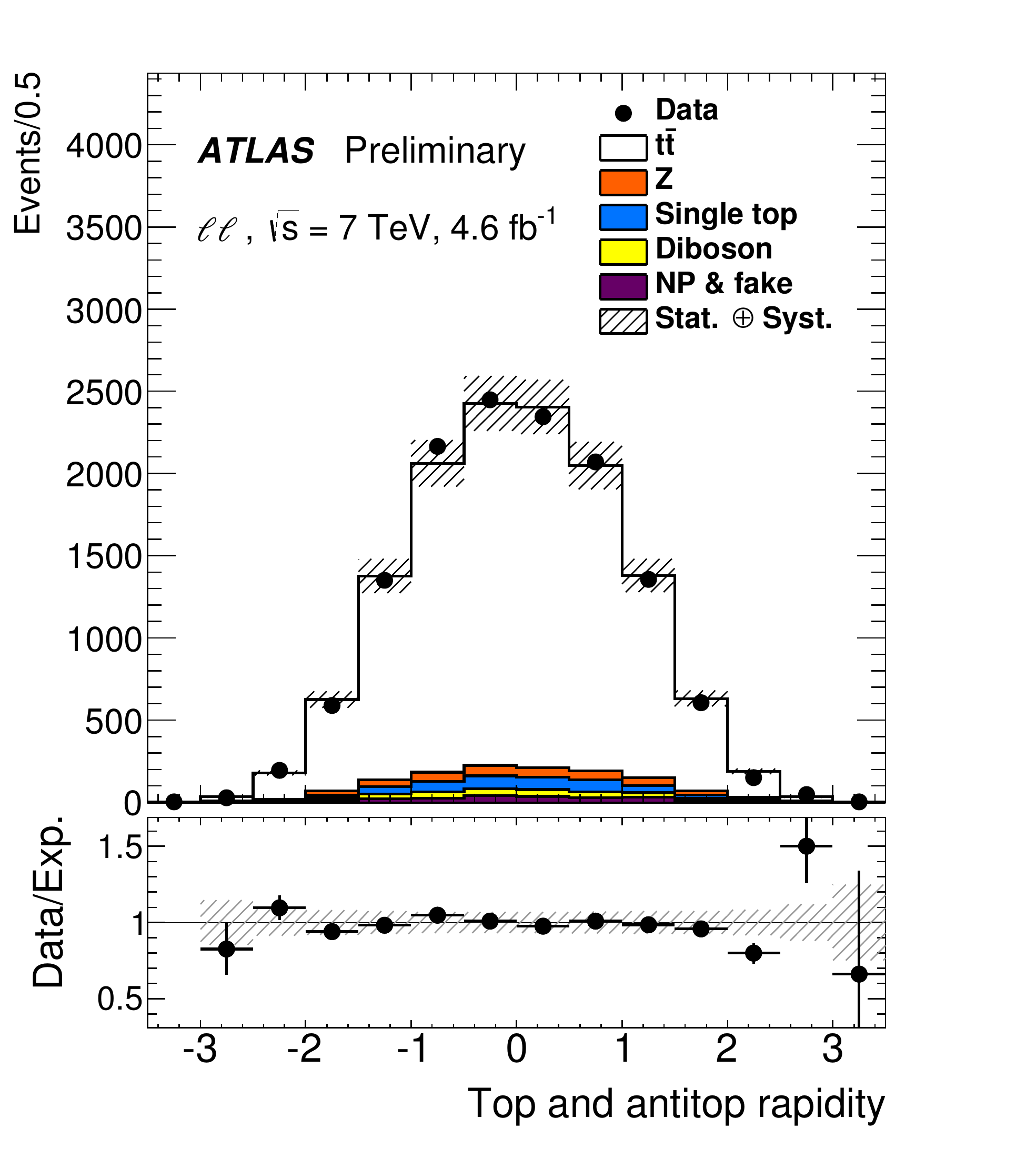}
}
\caption{\label{fig:dataMC_deta} Comparison of the expected and observed distributions of the (a) charged lepton $\eta${}  and (b) top and antitop rapidity distributions~\cite{asym-ll-7TeV}. The hatched area corresponds to the combined statistical and systematic uncertainties. "NP \& fake" refers to events with non-prompt or fake leptons.}
\end{figure}

Events are required to have been selected by a single-electron or single-muon trigger and to contain exactly two isolated, oppositely charged leptons. Depending on the lepton flavours, the sample is divided into three analysis channels referred to as \ee{}, \emu{} and \mumu{} in the following. Furthermore, the events are required to contain at least two jets with transverse momentum (\pt) above 25\,\GeV. In the \ee\ and \mumu\ channels, events are required to have the invariant mass of the two leptons $m_{\ell\ell} > 15$\,\GeV\ and $|m_{\ell\ell} - m_Z| > 10$\,\GeV\ and missing transverse momentum: $\met > 60$\,\GeV\ in order to reduce the  $Z/ \gamma^*$+jets production background. In the \emu{} channel, the smaller $Z/\gamma^*$+jets background is suppressed by requiring the scalar sum of the \pt{} of the jets and leptons ($\HT{}$) to be larger than 130\,\GeV. The background 
contributions are estimated using a data-driven technique for the $Z/\gamma^* \to ee/\mu\mu$ 
and non-prompt or fake leptons contributions. Other background contributions, including 
single-top production in the $Wt$ channel, $Z/ \gamma^* \to \tau\tau$ production and diboson events
($WW$, $WZ$ and $ZZ$) are estimated from Monte Carlo simulation. 
After the selection, the data sample contains about 8000 events, with an expected signal-to-background 
ratio of approximately six. The lepton $\eta$ and top rapidity distributions, which are the main inputs 
to the measurements, are modeled well by the simulation as shown in \FigRef{fig:dataMC_deta}.

\section{Kinematic reconstruction and corrections}
\label{Sec:RecoCorr}

The measurement of \Actt{} requires the reconstruction of the top and antitop kinematics. 
The neutrino weighting method~\cite{Abbott:1997fv} consisting of the following main steps is used. 
Assumptions on $\nu$ and $\nubar$ $\eta$ are made and the kinematic equations 
are solved for the top and antitop four-momenta. To each solution a weight is assigned, according to 
the compatibility between the measured \MET and the $\nu$ and $\nubar$ \pt. The solution 
corresponding to the maximum weight is selected to represent the event.

Both \Acll{} and \Actt{} measurements are corrected for the detector resolution and acceptance effects for the 
purpose of comparisons to state-of-the-art theory predictions. Due to the precise lepton reconstruction, the 
\Acll{} measurement requires small corrections of the \deta{} distribution, which are done using a bin-by-bin correction. 
On the other hand the \Actt{} measurement also uses the reconstructed jets and \MET{} and the top and antitop kinematics. The correspondingly larger corrections of the \dy{} distribution are done using the Fully Bayesian Unfolding technique~\cite{Fbu2012arXiv1201.4612C}.

\section{Results} 
\label{Sec:Results}
The differential cross-sections for \deta{} and \dy{} are obtained separately for each of the \ee, \emu{} and \mumu{} channels.  After the corrections, the results are combined using the Best Linear Unbiased Estimator (BLUE) method~\cite{Lyons:1988rp}. The inclusive corrected values are~\cite{asym-ll-7TeV}:
\begin{linenomath}
\begin{eqnarray}
 \Acll & = & \Acllcomb,
\end{eqnarray}
\end{linenomath}
\begin{linenomath}
\begin{eqnarray}
 \Actt & = & \Acttcomb.
\end{eqnarray}
\end{linenomath}
 For both the lepton-based asymmetry \Acll and the \ttbar{} asymmetry \Actt, the
statistical uncertainty is larger than the total systematic uncertainty. 
For the \Acll{} the largest contribution to the systematic uncertainty stems from the lepton reconstruction.
For the \Actt{} the largest sources of systematic uncertainty are from several sources of comparable size: the lepton reconstruction, the \met{} and the jet reconstruction uncertainty and the uncertainty due to the NP \& fake contribution. 

\begin{figure}
\centering
\subfigure[] {
\includegraphics[width=0.48\textwidth]{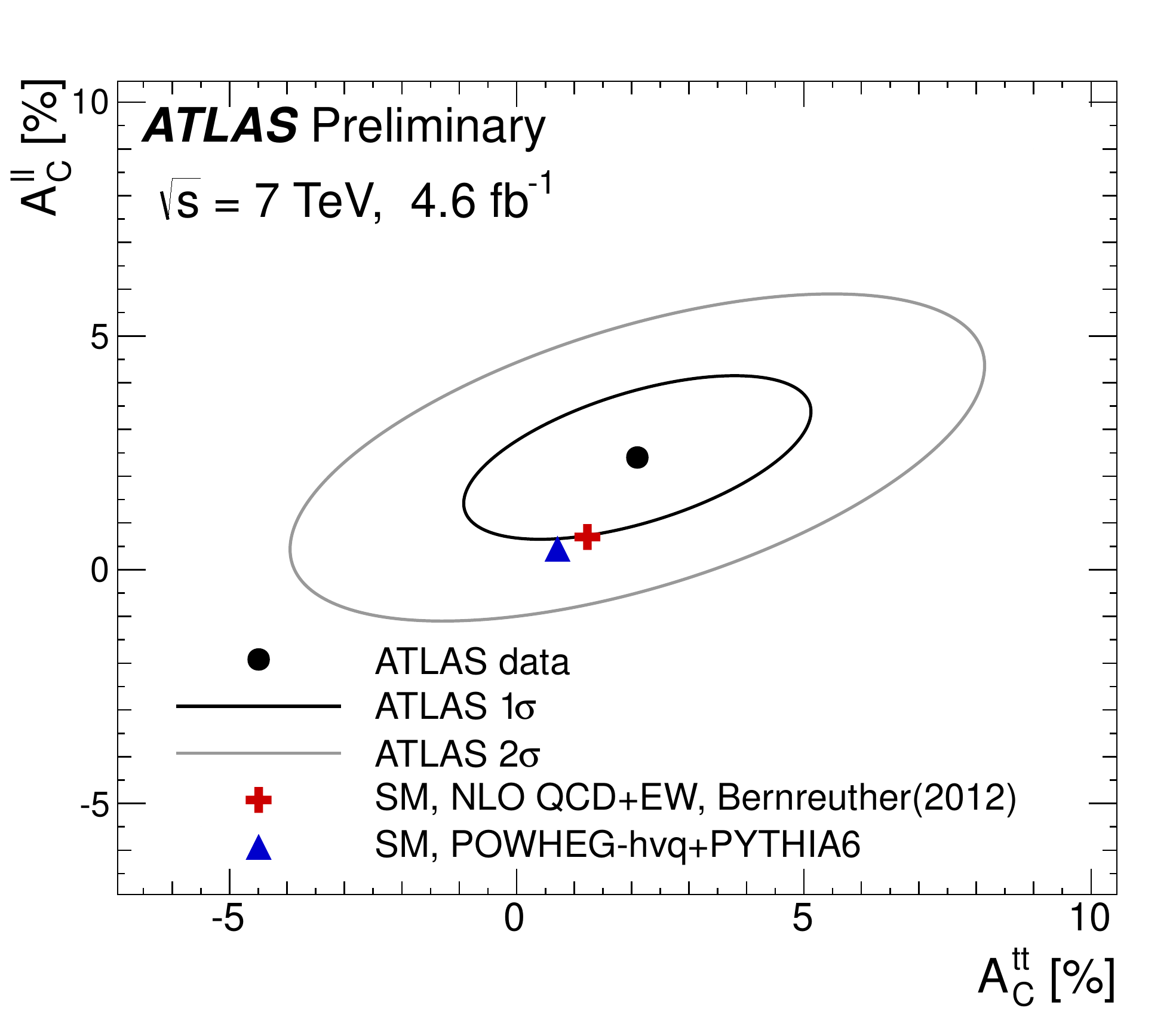}
\label{fig:acll_actt_twod_sm}
}
\subfigure[] {
\includegraphics[width=0.48\textwidth]{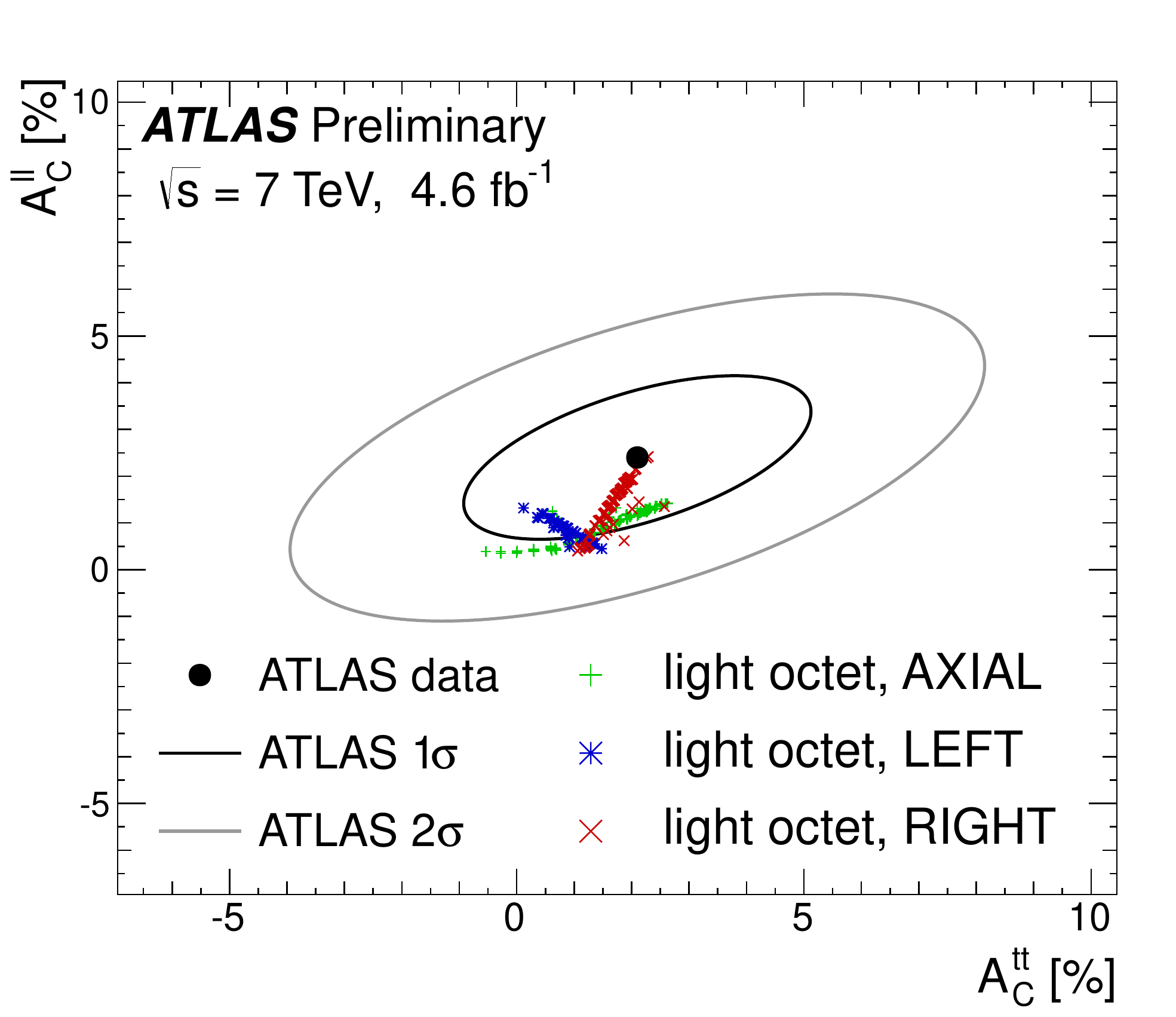}
\label{fig:acll_actt_twod_bsm}
}
\caption{\label{fig:acll_actt_twod} 
         Comparison of the inclusive \Acll{} and \Actt{} measurements to (a) SM theory predictions, 
         and (b) predictions of BSM models with a new colour octet particle~\cite{asym-ll-7TeV}.}
\end{figure}

In~\figRef{fig:acll_actt_twod} the measured values are compared to various theoretical predictions. The ellipses 
corresponding to 1$\sigma$ and 2$\sigma$ combined statistical and systematic uncertainties of the measurement, including the correlation between \Acll and \Actt are also shown. In \figRef{fig:acll_actt_twod_sm} the measured values are shown to be in good agreement with the state-of-the-art theory predictions obtained at NLO QCD including electromagnetic and weak-interaction corrections~\cite{Bernreuther:2012sx}. The predicted values are $\Acll = \Aclltheory$ and $\Actt = \Actttheory$, where the uncertainties are due to the renormalization and factorization scale variations. The predictions obtained by the \powheg~\cite{FRI-0701,Nason:2004rx,Frixione:2007nw}+\pythia~\cite{SJO-0601} generator, which is used for the signal sample simulation, are also shown. 
In~\figRef{fig:acll_actt_twod_bsm} the measurements are compared to the BSM models with a new colour octet particle exchanged in the s-channel with a mass of $m=250$\,\GeV{} and a width of $\Gamma=0.2m$. With a suitable parameter choice, these models improve the global fit to \ttbar{} observables at the Tevatron and the LHC, including the total cross-sections, various asymmetries, the top polarisation and spin correlations~\cite{Aguilar-Saavedra:2014nja}. The asymmetries used in the fit include Tevatron measurements and LHC measurements in the single-lepton \ttbar{} decay channel, while the LHC dilepton decay channel measurements are excluded from the fit. The considered couplings to the quarks are such that the global fit to \ttbar{} observables at the Tevatron and the LHC are consistent with the fitted measurements within two standard deviations. The predictions obtained with the left-handed, right-handed and axial coupling of the octet to the quarks are shown. While the models span a sizeable range of values in the \Acll{} and \Actt{} plane, their predictions are consistent with the measured values. 
It is thus not possible to exclude BSM contributions that are still allowed by previous Tevatron and LHC measurements.

\section{Conclusion} 
\label{Sec:Conclusion}
The measured inclusive \Acll and \Actt are found to be in good agreement with state-of-the-art SM predictions. A larger data set will enable more precise as well as more differential measurements of the asymmetry in the dileptonic decay channel, thus enhancing the requirements of the SM tests. Future \Acll{} and \Actt{} measurements with a larger data set could also further constrain the allowed couplings of the BSM models such as the colour-octet models.

\section*{Acknowledgments}
C~D acknowledges the support of the Helmholtz Association. L M acknowledges funding by the LabEx P2IO in the framework {\sl Investissements d'Avenir} (ANR-11-IDEX-0003-01) managed by the French National Research Agency (ANR) and funding by the European Research Council under the European Union’s Seventh Framework Programme ERC Grant Agreement n. 617185.

\section*{References}
\bibliography{proceedings}


\end{document}